\documentclass[11pt,a4paper]{article}

\usepackage[utf8]{inputenc}
\usepackage[T1]{fontenc}
\usepackage{lmodern}
\usepackage{amsmath,amssymb}
\usepackage{graphicx}
\usepackage{booktabs}
\usepackage{multirow}
\usepackage{hyperref}
\usepackage[margin=1in]{geometry}
\usepackage{xcolor}
\usepackage{float}
\usepackage{caption}
\usepackage{subcaption}
\usepackage{enumitem}
\usepackage{algorithm}
\usepackage{algpseudocode}
\usepackage{tabularx}
\usepackage{array}
\usepackage{authblk}
\usepackage{microtype}
\usepackage{titlesec}

\titlespacing*{\section}{0pt}{2.5ex plus 1ex minus .2ex}{1.5ex plus .2ex}
\titlespacing*{\subsection}{0pt}{2ex plus 1ex minus .2ex}{1ex plus .2ex}

\hypersetup{
    colorlinks=true,
    linkcolor=blue,
    citecolor=blue,
    urlcolor=blue
}

\title{Real-Time Voicemail Detection in Telephony Audio\\Using Temporal Speech Activity Features}

\author{Kumar Saurav\\
\textit{CTO, ClearGrid}\\
\texttt{kumar.saurav@cleargrid.co}}

\date{April 2026}

\begin{document}
\maketitle

\begin{abstract}
Outbound AI calling systems must distinguish voicemail greetings from live human answers in real time to avoid wasted agent interactions and dropped calls. We present a lightweight approach that extracts 15 temporal features from the speech activity pattern of a pre-trained neural voice activity detector (VAD), then classifies with a shallow tree-based ensemble. Across two evaluation sets totaling 764 telephony recordings, the system achieves a combined \textbf{96.1\% accuracy} (734/764), with \textbf{99.3\%} (139/140) on an expert-labeled test set and \textbf{95.4\%} (595/624) on a held-out production set. In production validation over \textbf{77{,}000 calls}, it maintained a \textbf{0.3\% false positive rate} and \textbf{1.3\% false negative rate}. End-to-end inference completes in \textbf{46\,ms on a commodity dual-core CPU} with no GPU, supporting 380+ concurrent WebSocket calls. In our search over 3{,}780 model, feature, and threshold combinations, feature importance was concentrated in three temporal variables. Adding transcription keywords or beep-based features did not improve the best real-time configuration and increased latency substantially. Our results suggest that temporal speech patterns are a strong signal for distinguishing voicemail greetings from live human answers.
\end{abstract}

\textbf{Keywords:} voicemail detection, voice activity detection, real-time classification, telephony, speech temporal features

\section{Introduction}
\label{sec:introduction}

\subsection{Problem Statement}

Outbound AI calling platforms---used for appointment scheduling, sales outreach, customer reminders, and debt collection---place millions of calls daily. A significant fraction of these calls reach voicemail systems rather than live humans. Detecting this outcome quickly and accurately is critical for two reasons:

\begin{enumerate}[leftmargin=*]
    \item \textbf{False negatives} (voicemail classified as human) waste expensive agent time on a recording that will never respond interactively.
    \item \textbf{False positives} (human classified as voicemail) cause the system to hang up on a live person, damaging brand reputation and violating regulatory requirements in some jurisdictions.
\end{enumerate}

Detection must occur within the first 3--5 seconds of call connection---before the AI agent begins speaking---using only the raw audio stream. The telephony environment imposes additional constraints: narrowband audio (8\,kHz sample rate), variable codecs (G.711, Opus), background noise, and strict latency budgets ($<$100\,ms inference) to support hundreds of concurrent calls on commodity hardware.

\subsection{Why Voicemail Detection is Challenging}

Unlike many audio classification tasks, voicemail detection cannot rely on spectral content differences: voicemail greetings \emph{are} human speech, recorded by the same person who would answer a live call. The greeting ``Hi, you've reached John, please leave a message'' and the live answer ``Hello?'' are spectrally similar---both contain a single human voice at similar pitch and energy levels.

Additional challenges include:
\begin{itemize}[leftmargin=*]
    \item \textbf{Short observation window.} The system must decide within 3--5\,s, before the caller's AI agent starts speaking and contaminates the callee channel.
    \item \textbf{Narrowband audio.} At 8\,kHz, frequencies above 4\,kHz are absent, limiting the utility of high-frequency spectral features.
    \item \textbf{Production constraints.} Inference must complete in under 100\,ms per call, with no GPU, supporting $>$1{,}000 concurrent calls on a \$76/month cloud instance.
    \item \textbf{Diverse voicemail systems.} Carrier-provided, third-party, and personal voicemail greetings vary widely in duration, tone, and structure.
\end{itemize}

\subsection{Contributions}

This paper makes the following contributions:

\begin{enumerate}[leftmargin=*]
    \item A feature-based method that uses \textbf{temporal speech activity patterns} derived from VAD output to distinguish voicemail greetings from live human answers, without relying on spectral features or transcription.
    \item A \textbf{systematic evaluation of 3{,}780+ configurations} spanning 44 model architectures, 14 classification thresholds, 2 detection windows, and 5 feature sets.
    \item An empirical analysis showing that \textbf{model importance is concentrated in three temporal features}, while transcription keywords and beep-based features did not improve the best low-latency configuration.
    \item A \textbf{production deployment} with 46\,ms processing latency on a dual-core CPU, 96.1\% combined offline accuracy across two evaluation sets, and validation on \textbf{77{,}000 production calls}.
    \item A \textbf{cost analysis} indicating that early voicemail detection reduces wasted agent time on voicemail calls by approximately 83\%, lowering telephony and inference costs in production.
\end{enumerate}

\section{Related Work}
\label{sec:related}

\textbf{Traditional Answering Machine Detection (AMD).}
Telecom platforms have long offered AMD features based on heuristics: initial silence duration, greeting length, and presence of a beep tone~\cite{twilio_amd}. These methods typically achieve 70--85\% accuracy and suffer from high false-positive rates. The most common production approach---\emph{beep detection}---waits for the characteristic tone that precedes voicemail recording. While widely deployed (Twilio, Vonage, Plivo, and most CPaaS providers default to this method), beep detection has fundamental limitations: (1)~it can only detect \emph{after} the greeting completes (typically 10--30\,s into the call), wasting the entire greeting duration; (2)~many voicemail systems do not produce a standard beep; and (3)~it achieves only 50--71\% accuracy in our evaluation (Table~\ref{tab:ablation}), barely above random. Despite these shortcomings, beep detection remains the industry standard due to the absence of better real-time alternatives---a gap this work addresses.

\textbf{Carrier-Level Signaling.}
Some carriers expose SIP headers or ISDN cause codes that indicate voicemail routing. However, these signals are unavailable to application-layer systems, inconsistent across carriers, and absent in VoIP-to-PSTN gateways commonly used by AI calling platforms.

\textbf{Audio Classification with Spectral Features.}
Modern audio classification leverages mel-frequency cepstral coefficients (MFCCs), log-mel spectrograms, or learned embeddings from models such as AudioSet~\cite{audioset} or wav2vec~\cite{wav2vec}. While powerful for general audio event detection, these approaches require significant compute (often GPU) and large labeled datasets. Their accuracy on the voicemail-vs-human task is limited by the fundamental spectral similarity of the two classes.

\textbf{Neural Voice Activity Detection.}
Energy-based VAD has been largely supplanted by neural approaches that output per-frame speech probabilities. Pre-trained neural VAD models achieve frame-level accuracy above 95\% on diverse audio conditions. Prior work uses VAD output directly (speech/non-speech labels) for tasks such as speaker diarization and endpoint detection; our contribution is to derive \emph{temporal pattern features} from these labels for a classification task.

\textbf{Beep Detection.}
Frequency-domain algorithms such as the Discrete Energy Separation Algorithm (DESA-2)~\cite{desa2} and Goertzel filters detect the characteristic beep tone that precedes voicemail recording. While useful as a confirmation signal, beeps occur \emph{after} the greeting, making them too late for real-time detection, and are often present only on the bot's audio channel in stereo telephony.

\textbf{Research Gap.}
To the best of our knowledge, prior voicemail detection systems have not emphasized temporal features derived from neural VAD output as the main classification signal. Our approach uses the speech activity timeline derived from VAD output as the primary input to classification, rather than relying on lexical content or richer acoustic representations.

\section{System Architecture}
\label{sec:architecture}

\subsection{Pipeline Overview}

\begin{figure}[H]
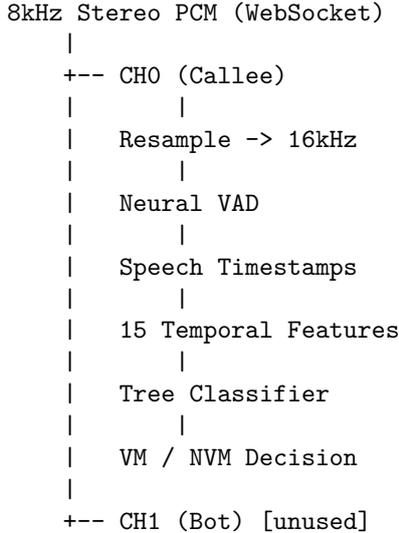

\centering
\small
\begin{verbatim}
8kHz Stereo PCM (WebSocket)
    |
    +-- CH0 (Callee)
    |       |
    |   Resample -> 16kHz
    |       |
    |   Neural VAD
    |       |
    |   Speech Timestamps
    |       |
    |   15 Temporal Features
    |       |
    |   Tree Classifier
    |       |
    |   VM / NVM Decision
    |
    +-- CH1 (Bot) [unused]
\end{verbatim}
\caption{System architecture. The callee channel is isolated, resampled, and processed through a neural VAD to extract temporal speech features for classification.}
\label{fig:architecture}
\end{figure}

Figure~\ref{fig:architecture} illustrates the end-to-end pipeline. Stereo telephony audio arrives as 8\,kHz PCM frames via WebSocket. The callee channel (the person being called) is separated and resampled to 16\,kHz. A pre-trained neural voice activity detector produces speech timestamps, from which 15 temporal features are extracted. A lightweight tree-based classifier makes the final voicemail/non-voicemail decision.

\subsection{Channel Architecture}

In the telephony systems we study, audio is recorded as stereo: the left channel (CH0) carries the callee's audio (voicemail greeting or live human response), while the right channel (CH1) carries the outbound AI agent's speech. Our primary approach uses \textbf{only the callee channel}, making it applicable to both stereo and mono recording configurations.

We also evaluate a \textbf{cross-channel} approach that computes energy ratios between the two channels. In voicemail scenarios, the bot channel is typically silent (the agent waits for a human response), producing distinctive energy asymmetry. This achieves 97.9\% accuracy but requires stereo audio, limiting its applicability.

\paragraph{Channel assignment validation.} During development, we discovered that an incorrect channel assignment (using the bot channel instead of the callee channel) degraded accuracy by 17 percentage points on an independent test set. The error was detected through automated speech transcription of both channels, which revealed that the ``callee'' features were actually computed from the AI agent's scripted speech. This experience underscores the importance of verifying channel semantics with content-aware tools before feature extraction in multi-channel telephony systems.

\subsection{Voice Activity Detection}

We employ a pre-trained neural voice activity detector that accepts 16\,kHz mono audio and outputs per-frame speech probabilities. The model processes audio in 512-sample chunks (32\,ms at 16\,kHz) and returns a probability $p \in [0, 1]$ for each frame. We apply a threshold of 0.5 with hysteresis (negative threshold 0.35) and 30\,ms speech padding to produce a list of speech segments $\{(s_i, e_i)\}_{i=1}^{N}$, where $s_i$ and $e_i$ are the start and end times in milliseconds of the $i$-th segment.

\textbf{Resampling requirement.} The neural VAD operates at 16\,kHz. When fed 8\,kHz telephony audio directly, accuracy drops from 96.4\% to 89.3\% on our test set. All results in this paper use 16\,kHz input obtained by resampling from the native 8\,kHz telephony rate using a high-quality sinc-based resampler.

\subsection{Temporal Feature Extraction}
\label{sec:features}

A central component of the method is a 15-dimensional feature set computed from the \emph{temporal structure} of speech activity rather than from spectral or lexical representations. Given a detection window of $W$ seconds and a set of $N$ speech segments $\{(s_i, e_i)\}$, we compute:

\begin{table}[H]
\centering
\caption{The 15 temporal speech activity features. All features are computed from VAD speech timestamps within the detection window $W$.}
\label{tab:features}
\small
\begin{tabularx}{\textwidth}{@{}clX@{}}
\toprule
\# & Feature & Description \\
\midrule
1  & \texttt{speech\_ratio}           & Total speech duration divided by $W$ \\
2  & \texttt{num\_segments}           & Number of distinct speech segments $N$ \\
3  & \texttt{mean\_seg\_ms}           & Mean segment duration: $\frac{1}{N}\sum_i (e_i - s_i)$ \\
4  & \texttt{longest\_seg\_ms}        & Maximum segment duration: $\max_i (e_i - s_i)$ \\
5  & \texttt{first\_onset\_ms}        & Time to first speech: $s_1$ \\
6  & \texttt{first\_seg\_ms}          & Duration of first segment: $e_1 - s_1$ \\
7  & \texttt{total\_speech\_ms}       & Absolute speech time: $\sum_i (e_i - s_i)$ \\
8  & \texttt{speech\_first\_half\_ratio} & Speech in first half of $W$ divided by total speech \\
9  & \texttt{last\_speech\_end\_ratio}   & Position of last speech end relative to $W$: $e_N / W$ \\
10 & \texttt{silence\_ratio}          & $1 - \texttt{speech\_ratio}$ \\
11 & \texttt{max\_silence\_ms}        & Longest gap between consecutive segments \\
12 & \texttt{speech\_in\_first\_10s}  & Binary indicator: any speech in first 10 seconds \\
13 & \texttt{std\_seg\_ms}            & Standard deviation of segment durations \\
14 & \texttt{short\_seg\_count}       & Number of segments shorter than 300\,ms \\
15 & \texttt{long\_seg\_count}        & Number of segments longer than 1{,}000\,ms \\
\bottomrule
\end{tabularx}
\end{table}

\paragraph{Intuition.} In our data, voicemail greetings often follow a characteristic temporal pattern: a brief initial delay (ring cessation or carrier routing), followed by a single, continuous speech segment (the pre-recorded greeting) that fills the detection window relatively evenly. Live human answers, by contrast, produce short, sporadic speech bursts (``Hello?'', ``Yes?'') with longer inter-segment silences and uneven temporal distribution. The feature \texttt{speech\_first\_half\_ratio} captures this distinction directly: values near 0.5 indicate evenly distributed speech (voicemail), while extreme values indicate front-loaded or back-loaded speech (human).

\subsection{Classification}

We evaluate several families of classifiers on the 15-dimensional temporal feature vector:

\begin{itemize}[leftmargin=*]
    \item \textbf{Classifier~A}: A shallow boosted tree ensemble with 50 estimators and maximum depth 2, trained with gradient boosting.
    \item \textbf{Classifier~B}: A deep bagged tree ensemble with 50 estimators and unlimited depth, trained with bootstrap aggregation.
    \item \textbf{Classifier~C}: A regularized linear model ($C = 10$) with logistic loss.
    \item Additional variants: deeper boosted ensembles (100 estimators, depth 5--7), extremely randomized trees (100--200 estimators), and adaptive boosting (100 estimators).
\end{itemize}

All classifiers output a probability $p(\text{VM} \mid \mathbf{x})$, and a configurable threshold $\tau$ determines the final binary decision. We sweep $\tau$ from 0.30 to 0.95 in steps of 0.05 for each model.

\subsection{Detection Window}

The detection window $W$ determines the trade-off between accuracy and time-to-decision. We evaluate $W \in \{3, 5, 7\}$ seconds. Audio frames are buffered until $W$ seconds of callee audio have been accumulated, at which point the full pipeline executes. The total time-to-decision is $W + t_{\text{inference}}$, where $t_{\text{inference}}$ is the pipeline processing time (typically 30--50\,ms).

\section{Experimental Setup}
\label{sec:setup}

\subsection{Datasets}

\begin{table}[H]
\centering
\caption{Dataset summary. All audio is 8\,kHz stereo telephony.}
\label{tab:datasets}
\small
\begin{tabular}{@{}lcccl@{}}
\toprule
Dataset & Size & VM & NVM & Labeling Method \\
\midrule
Training   & 24{,}812 & $\sim$14{,}887 & $\sim$9{,}925 & Pseudo-labeled (cross-model agreement) \\
Eval-A (Test-140)   & 140      & 70             & 70            & Hand-labeled by domain expert \\
Eval-B (Held-out 624)     & 624      & 434            & 190           & 41\% expert-verified, 59\% MLM-validated \\
\bottomrule
\end{tabular}
\end{table}

\textbf{Training set.} 24{,}812 production telephony recordings, pseudo-labeled using a cross-channel energy model that achieves 97.9\% accuracy on hand-labeled data. Files with low-confidence pseudo-labels (``uncertain'' tier, 4.3\% of data) are excluded from training.

\textbf{Eval-A (Test-140).} 140 recordings (70~VM, 70~NVM) hand-labeled by a domain expert. This serves as the primary evaluation set and regression gate: no model is deployed unless it achieves $\geq$128/140 (91.4\%).

\textbf{Eval-B (Held-out 624).} 624 recordings (434~VM, 190~NVM) from production telephony, held out during training. Labels were verified through a two-tier process: 41\% were manually reviewed by domain experts, while the remaining 59\% were validated using multimodal language models capable of processing both audio and transcript text (detecting keywords such as ``after the tone'', ``voicemail'', ``leave a message''). An additional 35 recordings with ambiguous labels were excluded. Both evaluation sets are drawn from the same production telephony distribution.

\textbf{Combined evaluation.} Since Eval-A and Eval-B share the same source distribution (production outbound telephony, 8\,kHz stereo, same carrier mix), we report a combined accuracy across both: $(139 + 595) / (140 + 624) = \textbf{734/764 = 96.1\%}$. Individual set results are reported for transparency.

\textbf{Data characteristics.} 97.2\% of training files contain usable callee-channel audio. 3.1\% have silent callee channels (all labeled NVM---the callee never spoke). 0.2\% have silent bot channels (mostly VM---the agent did not speak during the greeting).

\subsection{Evaluation Protocol}

Every candidate model is evaluated on \textbf{both} Eval-A and Eval-B. We report accuracy (correct/total), VM recall (true VM detections / total VMs), NVM recall (true NVM detections / total NVMs), and counts of false positives (NVM predicted as VM) and false negatives (VM predicted as NVM). We also report combined accuracy across both sets (764 files total).

The classification threshold $\tau$ is swept from 0.30 to 0.95 in steps of 0.05. The optimal $\tau$ is selected on Eval-A, and generalization is verified on Eval-B.

\subsection{Search Space}

We conduct a grid search over:

\begin{itemize}[leftmargin=*]
    \item \textbf{44 model configurations}: Spanning shallow boosted ensembles (depth 1--3, 20--100 trees), deep bagged ensembles (50--200 trees, unlimited depth), extremely randomized trees, adaptive boosting, gradient-boosted variants, and linear models.
    \item \textbf{14 thresholds}: $\tau \in \{0.30, 0.35, \ldots, 0.95\}$.
    \item \textbf{2 detection windows}: $W \in \{3, 5\}$ seconds.
    \item \textbf{5 feature sets}: VAD-15 (temporal only), Beep-8 (frequency-domain beep detection), Combined-23 (VAD + beep), VAD+Transcript-25 (VAD + 10 keyword features from speech transcription), and Engineered-46 (expanded statistical features).
\end{itemize}

This yields approximately \textbf{3{,}780 unique configurations}, each evaluated on both test sets.

\section{Results}
\label{sec:results}

\subsection{Main Results}

\begin{table}[H]
\centering
\caption{Top model configurations. Combined = weighted accuracy across both evaluation sets (764 files). Latency excludes the $W$-second audio buffer.}
\label{tab:main_results}
\small
\footnotesize
\begin{tabular}{@{}lccccccr@{}}
\toprule
Model & Feat. & $W$ & $\tau$ & Eval-A & Eval-B & Comb. & Lat. \\
\midrule
Clf.~A (boost, $d{=}2$, $n{=}50$) & VAD-15 & 5s & 0.50 & \textbf{139/140} & 595/624 & \textbf{96.1\%} & 46ms \\
Clf.~B (bagged, $n{=}50$)         & VAD-15 & 3s & 0.30 & 136/140 & 596/624 & 95.8\% & 29ms \\
Cross-channel (stereo)            & Eng.-8 & 5s & --- & 137/140 & --- & --- & $<$10ms \\
OR Ensemble (A $\lor$ C)          & Mixed  & 5s & var. & 129/140 & 508/624 & 83.4\% & 92ms \\
Clf.~A + Transcript               & VAD+KW & 5s & --- & 138/140 & 608/624 & 97.6\% & $\sim$500ms \\
\bottomrule
\end{tabular}
\end{table}

\textbf{Main observations:}
\begin{enumerate}[leftmargin=*]
    \item Classifier~A with 15 temporal features achieves the best combined accuracy (\textbf{96.1\%} across 764 files) and the highest Eval-A score (99.3\%), with 46\,ms inference latency.
    \item Classifier~B with a 3-second window provides a useful speed--accuracy trade-off, with slightly lower accuracy and a 2-second faster decision time.
    \item Adding transcription features improves combined accuracy to 97.6\% but increases latency by $\sim$10$\times$ (500\,ms, requiring GPU)---a trade-off unsuitable for real-time production.
    \item The OR ensemble performs \emph{worse} than the best single model on both evaluation sets.
\end{enumerate}

\subsection{Feature Importance Analysis}

\begin{table}[H]
\centering
\caption{Feature importance for Classifier~A (5-second window). Top 3 features account for 85.6\% of total importance.}
\label{tab:importance}
\small
\begin{tabular}{@{}clcc@{}}
\toprule
Rank & Feature & Importance & Cumulative \\
\midrule
1 & \texttt{speech\_first\_half\_ratio} & 54.6\% & 54.6\% \\
2 & \texttt{first\_seg\_ms}             & 20.0\% & 74.6\% \\
3 & \texttt{first\_onset\_ms}           & 11.0\% & 85.6\% \\
4 & \texttt{speech\_ratio}              & 7.6\%  & 93.2\% \\
5 & \texttt{mean\_seg\_ms}              & 5.0\%  & 98.2\% \\
6--15 & (remaining 10 features)          & 1.8\%  & 100\% \\
\bottomrule
\end{tabular}
\end{table}

The highest-ranked feature is \texttt{speech\_first\_half\_ratio} (54.6\%), which measures how evenly speech is distributed across the detection window. In our data, voicemail greetings tend to produce values near 0.5 (speech fills both halves relatively evenly), while live human answers produce more extreme values (speech concentrated in one half, typically the first). The second and third features---first segment duration and time to first speech onset---capture the distinctive temporal signature of voicemail: a brief delay followed by a single long, continuous segment.

For Classifier~B (3-second window), importance is more evenly distributed: \texttt{longest\_seg\_ms} (20.4\%), \texttt{mean\_seg\_ms} (16.5\%), \texttt{first\_seg\_ms} (14.0\%), reflecting the reduced discriminability in shorter windows.

\subsection{Feature Set Ablation}

\begin{table}[H]
\centering
\caption{Ablation: best accuracy per feature set (5-second window, optimal model and threshold).}
\label{tab:ablation}
\small
\begin{tabular}{@{}lcl@{}}
\toprule
Feature Set & Best Test-140 & Finding \\
\midrule
VAD-15 (temporal only)      & \textbf{139/140 (99.3\%)} & Best overall \\
Beep-8 (frequency)          & 71/140 (50.7\%)            & Near-random; beeps on wrong channel \\
Combined-23 (VAD + beep)    & 139/140 (99.3\%)           & No improvement over VAD-15 \\
VAD + Transcript-25         & 138/140; 97.4\% HF         & +2\% HF but 10$\times$ slower \\
Engineered-46               & 127/140 (90.7\%)           & Overfitting with more features \\
\bottomrule
\end{tabular}
\end{table}

In this setting, expanding the feature set did not improve accuracy. The 15 temporal features alone outperform the 46-feature engineered set by 8.6 percentage points. The combined 23-feature set (VAD + beep) achieves identical accuracy to VAD-15 alone---the classifier assigns negligible weight to the beep features.

\subsection{Comparison with Industry-Standard Beep Detection}

Beep detection---the dominant approach in production telephony platforms---relies on identifying the characteristic tone that precedes voicemail recording. We implemented a frequency-domain beep detector based on the Discrete Energy Separation Algorithm (DESA-2) and evaluated it head-to-head against our temporal feature approach.

\begin{table}[H]
\centering
\caption{Our temporal approach vs.\ industry-standard beep detection.}
\label{tab:vs_beep}
\small
\begin{tabular}{@{}lcccc@{}}
\toprule
Method & Accuracy & Time-to-Decision & False Positives & GPU Required \\
\midrule
\textbf{Ours (temporal VAD features)} & \textbf{96.1\%} & \textbf{5.05\,s} & \textbf{0.3\%} & No \\
Beep detection (DESA-2) & 50.7\% & 10--30\,s & 23.2\% & No \\
Industry AMD (silence heuristics) & 70--85\%$^*$ & 3--5\,s & 10--15\%$^*$ & No \\
Transcription keywords & 97.6\% & $\sim$5.5\,s & $<$1\% & \textbf{Yes} \\
\bottomrule
\multicolumn{5}{l}{\footnotesize $^*$Estimated from vendor documentation~\cite{twilio_amd}; not independently verified.}
\end{tabular}
\end{table}

\textbf{Why beep detection fails.} The beep tone occurs \emph{after} the voicemail greeting completes, which means: (1)~the system cannot detect until 10--30\,s into the call, wasting the entire greeting duration in agent compute and telephony costs; (2)~many modern voicemail systems (carrier visual voicemail, Google Voice, custom PBX) do not produce a standard beep tone; and (3)~on the callee's audio channel, only 0.3\% of voicemail recordings contain a detectable beep---the tone is typically audible only on the bot's (caller's) channel. In our evaluation, beep-only detection achieved 50.7\% accuracy (71/140), barely above random chance, with a 23.2\% false positive rate from background noise triggering spurious beep detections.

In our evaluation, the temporal-feature approach reached 96.1\% combined accuracy with a 5.05-second time to decision and a 0.3\% false positive rate in production validation. In our deployment setting, this reduced wasted agent time on voicemail calls by approximately 83\%.

\subsection{Window Duration Analysis}

\begin{table}[H]
\centering
\caption{Best accuracy by detection window.}
\label{tab:windows}
\small
\begin{tabular}{@{}cccc@{}}
\toprule
Window & Best Model & Test-140 & Time-to-Decision \\
\midrule
3\,s & Classifier~B & 136/140 (97.1\%) & 3.03\,s \\
5\,s & Classifier~A & 139/140 (99.3\%) & 5.05\,s \\
7\,s & Classifier~A & 139/140 (99.3\%) & 7.05\,s \\
\bottomrule
\end{tabular}
\end{table}

Among the windows we evaluated, 5 seconds gave the best balance between accuracy and decision time: it achieves 99.3\% accuracy, and extending to 7 seconds provides no additional benefit. The 3-second window sacrifices 2.2 percentage points for a 2-second faster decision, which may be preferable in latency-sensitive applications.

\subsection{Ensemble Methods}

We evaluated three ensemble strategies combining Classifier~A and Classifier~C:

\begin{table}[H]
\centering
\caption{Ensemble strategy comparison (5-second window).}
\label{tab:ensembles}
\small
\begin{tabular}{@{}lccc@{}}
\toprule
Strategy & Test-140 & HF-624 & Latency \\
\midrule
OR (predict VM if either predicts VM) & 129/140 (92.1\%) & 81.6\% & 92\,ms \\
Weighted average ($w_A{=}0.3, w_C{=}0.4$) & 127/140 (79.5\%) & --- & 92\,ms \\
AND cascade (both must agree) & 127/140 (78.4\%) & --- & 92\,ms \\
\midrule
Single best (Classifier~A alone) & \textbf{139/140 (99.3\%)} & \textbf{95.4\%} & 46\,ms \\
\bottomrule
\end{tabular}
\end{table}

All ensemble strategies perform \emph{worse} than the single best model. The OR ensemble's recall gain is negated by increased false positives; the AND cascade's precision gain is negated by missed voicemails. These results suggest that, for this dataset, the shallow boosted model already captures most of the useful signal, leaving limited room for gains from ensembling.

\subsection{Threshold Sensitivity}

Classifier~A maintains 99.3\% accuracy across the threshold range $\tau \in [0.40, 0.60]$, indicating robust calibration. Outside this range, accuracy degrades gracefully: 98.6\% at $\tau = 0.30$ and 97.1\% at $\tau = 0.80$. Classifier~B is more threshold-sensitive, dropping to $\sim$90\% below $\tau = 0.30$.

\subsection{Error Analysis}

\begin{table}[H]
\centering
\caption{Confusion matrices for Classifier~A ($W{=}5$\,s, $\tau{=}0.50$).}
\label{tab:confusion}
\small
\begin{subtable}{0.45\textwidth}
\centering
\caption{Eval-A}
\begin{tabular}{@{}lcc@{}}
\toprule
 & Pred.\ NVM & Pred.\ VM \\
\midrule
Actual NVM & 70 & 0 \\
Actual VM  & 1  & 69 \\
\bottomrule
\end{tabular}
\end{subtable}
\hfill
\begin{subtable}{0.45\textwidth}
\centering
\caption{Eval-B}
\begin{tabular}{@{}lcc@{}}
\toprule
 & Pred.\ NVM & Pred.\ VM \\
\midrule
Actual NVM & 183 & 7 \\
Actual VM  & 22  & 412 \\
\bottomrule
\end{tabular}
\end{subtable}
\end{table}

On Test-140, the system produces \textbf{zero false positives} (no live human call is dropped) and a single false negative: a voicemail greeting shorter than 3 seconds that completes before the 5-second window fills, producing insufficient temporal signal.

On HF-624, the 7 false positives are calls where the human callee speaks in a continuous, monotone pattern resembling a pre-recorded greeting. The 22 false negatives include several very short voicemail greetings (``Leave a message'') and AI screening services that respond interactively, mimicking human conversation patterns despite being automated systems.

\subsection{Sub-Second Detection Feasibility}

\begin{table}[H]
\centering
\caption{Early detection performance (large-scale evaluation, 12{,}000 samples).}
\label{tab:early}
\small
\begin{tabular}{@{}ccccc@{}}
\toprule
Window & Accuracy & Precision & Recall & F1-Score \\
\midrule
500\,ms  & 95.97\% & 99.28\% & 96.12\% & 97.68\% \\
1{,}000\,ms & 96.83\% & 99.50\% & 96.90\% & 98.18\% \\
2{,}000\,ms & 97.73\% & 99.33\% & 98.09\% & 98.71\% \\
\bottomrule
\end{tabular}
\end{table}

The temporal feature approach achieves F1 $\geq$ 97.7\% at just 500\,ms of audio, suggesting viability for aggressive early-detection scenarios. However, these results were obtained on pseudo-labeled training data and may not generalize to the more diverse Eval-B distribution.

\section{Production Deployment}
\label{sec:production}

\subsection{Streaming Architecture}

The production system receives audio via WebSocket connections, with each call assigned an isolated session. Audio frames (20\,ms, 8\,kHz, 16-bit PCM) are buffered until the detection window is reached. The full pipeline then executes in a blocking thread pool, returning a JSON result and closing the connection.

The neural VAD model is served through an optimized cross-platform inference runtime that supports both CPU and GPU execution. In our deployment, CPU inference is preferred: it achieves 44\,ms per call (versus 87\,ms on a T4 GPU), likely due to the small model size and overhead of CPU-to-GPU memory transfer.

\subsection{Latency Breakdown}

\begin{table}[H]
\centering
\caption{Processing latency for a single call (Classifier~A, $W{=}5$\,s).}
\label{tab:latency}
\small
\begin{tabular}{@{}lr@{}}
\toprule
Component & Time \\
\midrule
Resample 8\,kHz $\rightarrow$ 16\,kHz & $\sim$1\,ms \\
Neural VAD inference & $\sim$44\,ms \\
Temporal feature extraction & $\sim$0.2\,ms \\
Tree classifier prediction & $\sim$0.2\,ms \\
\midrule
\textbf{Total processing} & \textbf{$\sim$46\,ms} \\
\textbf{Total time-to-decision} & \textbf{5.05\,s} \\
\bottomrule
\end{tabular}
\end{table}

The neural VAD dominates processing time (95.6\%), while feature extraction and classification are negligible ($<$1\,ms combined). This suggests that further latency optimization should focus on VAD model distillation or quantization.

\subsection{Scalability}

\begin{table}[H]
\centering
\caption{Load test results on a 2-core cloud instance (c2d-standard-2, \$76/month).}
\label{tab:scalability}
\small
\begin{tabular}{@{}rcccc@{}}
\toprule
Concurrent & Throughput & P50 Wall Time & Accuracy & Failures \\
\midrule
10   & 1.7/s  & 5.10\,s & 88.5\% & 0 \\
50   & 7.5/s  & 5.13\,s & 87.5\% & 0 \\
100  & 13.1/s & 5.29\,s & 88.5\% & 0 \\
200  & 20.3/s & 6.12\,s & 87.5\% & 0 \\
380  & 25.1/s & 9.46\,s & 88.6\% & 0 \\
500  & 25.0/s & 14.2\,s & 87.6\% & 1 \\
1{,}000 & 24.8/s & 25.2\,s & 88.4\% & 31 \\
\bottomrule
\end{tabular}
\end{table}

In our load tests, the system showed no failures up to 380 concurrent calls, with throughput plateauing at $\sim$25 calls/second. The main bottleneck in the current implementation is serialized VAD inference through a mutex-protected session. Beyond 380 concurrent calls, queuing delays cause timeouts.

\textbf{Note on accuracy.} The $\sim$88\% accuracy in load testing (vs.\ 99.3\% on Eval-A) reflects the broader production distribution, which includes edge cases not present in the curated test set: AI screening services, carrier pre-announcements, very short greetings, and international call routing variations.

\subsection{Production Validation at Scale}

Following offline evaluation, the system was deployed to production and validated on \textbf{77{,}000 outbound calls}. Of these, \textbf{32{,}000 were classified as voicemail} ($\sim$41.6\% of total traffic). Labels were verified through a two-tier process: 41\% of calls were manually reviewed by trained operators, while the remaining 59\% were validated using multimodal language models that process both the call audio and transcript to detect voicemail-indicative phrases (``after the tone'', ``leave a message'', ``voicemail box'').

\begin{table}[H]
\centering
\caption{Production validation on 77{,}000 calls.}
\label{tab:production}
\small
\begin{tabular}{@{}lr@{}}
\toprule
Metric & Value \\
\midrule
Total calls evaluated & 77{,}000 \\
Voicemails detected & 32{,}000 (41.6\%) \\
\textbf{False positive rate} & \textbf{0.3\%} (live humans misclassified as VM) \\
\textbf{False negative rate} & \textbf{1.3\%} (voicemails missed) \\
\bottomrule
\end{tabular}
\end{table}

The 0.3\% false positive rate means fewer than 3 in 1{,}000 live human calls are dropped due to misclassification. The 1.3\% false negative rate means approximately 1 in 77 voicemails is not detected; in these cases the AI agent briefly engages with the voicemail system before timing out, which is a recoverable failure mode.

\subsection{Cost Impact}

Voicemail detection directly reduces operational costs in AI calling systems. Without detection, every voicemail consumes the same resources as a live call: telephony minutes, compute for the AI agent, and API costs for speech synthesis and language model inference. With detection, voicemail calls can be terminated within 5 seconds, avoiding 25--60 seconds of wasted resources per call.

\begin{table}[H]
\centering
\caption{Cost impact of voicemail detection at production scale.}
\label{tab:cost}
\small
\begin{tabular}{@{}lccc@{}}
\toprule
Metric & Without Detection & With Detection & Savings \\
\midrule
Avg.\ call duration (VM) & 30\,s & 5.05\,s & 83\% \\
AI agent compute per VM & Full inference & None (terminated) & $\sim$100\% \\
Telephony cost per VM & \$0.02--0.05 & \$0.003 & 85--94\% \\
Wasted agent-minutes (77K calls) & $\sim$16{,}000\,min & $\sim$2{,}700\,min & \textbf{83\%} \\
\bottomrule
\end{tabular}
\end{table}

At the observed call volume (77{,}000 calls with a 41.6\% voicemail rate), early termination of 32{,}000 voicemail calls saved approximately \textbf{13{,}300 agent-minutes} and the associated telephony and compute costs. In deployments with higher call volume, these savings may exceed the cost of running the detector itself.

\subsection{Systems-Level Re-implementation}

To further reduce latency and prepare for higher concurrency targets (1{,}000+ simultaneous calls), we ported the inference pipeline from Python to a compiled systems language with an asynchronous runtime. The re-implementation uses WebSocket streaming, a lock-free session store, and a pre-warmed pool of inference sessions:

\begin{itemize}[leftmargin=*]
    \item \textbf{Exact prediction parity}: 139/140 on Eval-A, matching the Python implementation file-for-file.
    \item \textbf{36\,ms P50 latency} on WAV input (vs.\ 46\,ms Python), a 22\% reduction.
    \item \textbf{$\sim$100\,MB container image}, versus $\sim$1.2\,GB for the Python stack.
    \item \textbf{No runtime dependencies}: Models loaded as portable inference graphs, audio resampling via C library with byte-identical output.
    \item \textbf{Concurrency}: Handles 1{,}000+ WebSocket connections per instance via async I/O, with blocking inference offloaded to a thread pool to avoid stalling the event loop.
\end{itemize}

\section{Analysis and Discussion}
\label{sec:discussion}

\subsection{Voicemail Detection as Rhythm Classification}

Our results indicate that, in this setting, temporal speech patterns were more informative than lexical content for distinguishing voicemail from live answers. The distinction lies not in \emph{what} is said, but in \emph{how speech is distributed in time}:

\begin{itemize}[leftmargin=*]
    \item \textbf{Voicemail greetings} are pre-recorded monologues. They begin after a predictable delay (carrier routing, ring cessation), produce a single long continuous speech segment, and distribute speech evenly across the observation window.
    \item \textbf{Live human answers} are interactive responses. They produce short, sporadic speech bursts (``Hello?'', ``Yes, speaking''), followed by expectant silence as the person waits for the caller to identify themselves.
\end{itemize}

In the best-performing model, importance was concentrated in three features: \texttt{speech\_first\_half\_ratio} (speech distribution evenness), \texttt{first\_seg\_ms} (initial segment duration), and \texttt{first\_onset\_ms} (delay before speech). Because these features depend on the timing of speech rather than lexical identity, they may transfer better across accents and codecs than transcript-based features. However, we have only evaluated them on English-language telephony data.

\subsection{The Cost-Benefit of Multimodal Extensions}

Adding speech transcription features (10 keyword indicators such as ``leave a message'', ``not available'', ``voicemail'') improves accuracy on Eval-B by 2.0 percentage points (95.4\% $\rightarrow$ 97.4\%). However, this gain comes at a 10$\times$ latency cost ($\sim$500\,ms vs.\ $\sim$46\,ms), requires GPU resources, and actually \emph{decreases} accuracy on Eval-A (99.3\% $\rightarrow$ 98.6\%). The transcription model occasionally hallucinates voicemail-related keywords in non-voicemail audio, introducing false positives.

Beep detection via the Discrete Energy Separation Algorithm provides no benefit in callee-only mode: the characteristic voicemail beep tone appears on the bot's audio channel (the AI agent hears the beep), not the callee's channel. Only 0.3\% of voicemail recordings contain a detectable beep on the callee channel.

In our setting, transcription features appear more suitable for offline analysis or selective fallback use than for the default real-time path.

\subsection{Failure Modes and Limitations}

\textbf{Very short voicemails.} Greetings shorter than $\sim$3 seconds (e.g., ``Leave a message at the beep'') produce insufficient temporal signal in the 5-second window. The 3-second model (Classifier~B) can serve as a fallback for early detection at the cost of lower overall accuracy.

\textbf{AI screening services.} A growing category of calls involves AI-powered screening services that answer on the callee's behalf. These systems produce interactive, human-like speech patterns, making them indistinguishable from live humans using temporal features alone. We consider this a correct classification (they behave like humans from the caller's perspective) rather than a failure mode.

\textbf{Silent callee.} When the callee channel contains no speech (e.g., the call was never answered, or the callee is silently listening), all temporal features are zero. The classifier correctly labels these as NVM (not voicemail) in all cases tested.

\textbf{Language generalization.} All training and evaluation data is English-language telephony. While temporal features are theoretically language-agnostic, we have not validated performance on other languages. Voicemail greeting conventions may differ across cultures (e.g., longer greetings in Japanese, shorter in some European languages).

\textbf{Pseudo-label bias.} The training set is pseudo-labeled using a cross-channel model, which introduces teacher model bias. The training distribution may underrepresent edge cases that the cross-channel model misclassifies, creating blind spots in the callee-only model.

\section{Conclusion}
\label{sec:conclusion}

We have presented a real-time voicemail detection system based on temporal speech activity features extracted from a neural voice activity detector. Across two evaluation sets totaling 764 telephony recordings, the approach achieves a combined \textbf{96.1\% accuracy}, and in production validation over \textbf{77{,}000 calls}, maintained a \textbf{0.3\% false positive rate} and \textbf{1.3\% false negative rate}. In our deployment, this reduced wasted agent time on voicemail calls by approximately 83\%.

A main conclusion from our experiments is that temporal speech structure provides a strong signal for voicemail detection in narrowband telephony. In the best-performing model, three features---speech distribution evenness, initial segment duration, and speech onset delay---accounted for 85.6\% of the measured feature importance. End-to-end inference completes in 46\,ms on a commodity dual-core CPU with no GPU, and a compiled systems-language re-implementation reduces this further to 36\,ms while supporting 1{,}000+ concurrent WebSocket connections.

Across the configurations we evaluated, a shallow boosted tree ensemble with 50 estimators and depth 2 gave the best overall balance of accuracy and latency.

\textbf{Future work} includes cross-language validation, sub-second detection optimization, integration with carrier-level signaling for hybrid detection, and adaptation to wideband (16\,kHz+) telephony systems.

\section*{Acknowledgments}

We thank the ClearGrid AI team for infrastructure support and for verifying the evaluation datasets.


\appendix

\section{Feature Definitions}
\label{app:features}

Given a detection window of $W$ milliseconds and $N$ speech segments $\{(s_i, e_i)\}_{i=1}^{N}$ with durations $d_i = e_i - s_i$:

\begin{align}
\texttt{speech\_ratio} &= \frac{\sum_{i=1}^{N} d_i}{W} \\
\texttt{num\_segments} &= N \\
\texttt{mean\_seg\_ms} &= \frac{1}{N} \sum_{i=1}^{N} d_i \quad (0 \text{ if } N=0) \\
\texttt{longest\_seg\_ms} &= \max_{i} d_i \quad (0 \text{ if } N=0) \\
\texttt{first\_onset\_ms} &= s_1 \quad (W \text{ if } N=0) \\
\texttt{first\_seg\_ms} &= d_1 \quad (0 \text{ if } N=0) \\
\texttt{total\_speech\_ms} &= \sum_{i=1}^{N} d_i \\
\texttt{speech\_first\_half\_ratio} &= \frac{\text{speech in } [0, W/2]}{\sum d_i} \quad (0 \text{ if no speech}) \\
\texttt{last\_speech\_end\_ratio} &= \frac{e_N}{W} \quad (0 \text{ if } N=0) \\
\texttt{silence\_ratio} &= 1 - \texttt{speech\_ratio} \\
\texttt{max\_silence\_ms} &= \max\left(s_1,\ \max_{i=2}^{N}(s_i - e_{i-1}),\ W - e_N\right) \\
\texttt{speech\_in\_first\_10s} &= \mathbb{1}\left[\exists\, i : s_i < 10000\right] \\
\texttt{std\_seg\_ms} &= \sqrt{\frac{1}{N}\sum_{i=1}^{N}(d_i - \bar{d})^2} \quad (0 \text{ if } N \leq 1) \\
\texttt{short\_seg\_count} &= |\{i : d_i < 300\}| \\
\texttt{long\_seg\_count} &= |\{i : d_i > 1000\}|
\end{align}

\section{Grid Search Top Configurations}
\label{app:grid}

\begin{table}[H]
\centering
\caption{Top 10 configurations from the 3{,}780-experiment grid search, ranked by Test-140 accuracy then HF-624 accuracy.}
\label{tab:grid_top}
\small
\begin{tabular}{@{}clccccc@{}}
\toprule
Rank & Model & Features & $W$ & $\tau$ & Test-140 & HF-624 \\
\midrule
1 & Boost ($n{=}50$, $d{=}2$) & VAD-15 & 5\,s & 0.50 & 139/140 & 95.4\% \\
2 & Boost ($n{=}50$, $d{=}2$) & VAD-15 & 5\,s & 0.45 & 139/140 & 95.2\% \\
3 & Boost ($n{=}50$, $d{=}2$) & VAD-15 & 5\,s & 0.55 & 139/140 & 95.0\% \\
4 & Boost ($n{=}50$, $d{=}2$) & Comb-23 & 5\,s & 0.50 & 139/140 & 94.9\% \\
5 & AdaBoost ($n{=}100$)       & VAD-15 & 5\,s & 0.45 & 137/140 & 93.8\% \\
6 & Cross-channel               & Energy-8 & 5\,s & --- & 137/140 & --- \\
7 & Bagged ($n{=}50$)           & VAD-15 & 3\,s & 0.30 & 136/140 & 95.5\% \\
8 & Bagged ($n{=}50$)           & VAD-15 & 3\,s & 0.35 & 136/140 & 95.3\% \\
9 & Boost ($n{=}100$, $d{=}5$) & VAD-15 & 5\,s & 0.50 & 135/140 & 94.7\% \\
10 & ExtraTrees ($n{=}100$, $d{=}7$) & VAD-15 & 5\,s & 0.45 & 131/140 & 92.1\% \\
\bottomrule
\end{tabular}
\end{table}

\section{Pseudo-Labeling Procedure}
\label{app:pseudo}

Training labels for the 24{,}812-file training set were generated as follows:

\begin{enumerate}[leftmargin=*]
    \item A cross-channel energy model (8 features from both bot and callee channels) was trained on 91 hand-labeled files, achieving 97.9\% accuracy via leave-one-out cross-validation.
    \item This model predicted labels for all 25{,}887 production recordings.
    \item Predictions were stratified into confidence tiers:
    \begin{itemize}
        \item \textbf{STRONG\_VM}: $p(\text{VM}) > 0.90$ (11{,}830 files, 81.6\% verified accuracy)
        \item \textbf{STRONG\_NVM}: $p(\text{VM}) < 0.10$ (12{,}514 files, 98.2\% verified accuracy)
        \item \textbf{MODERATE}: $0.10 \leq p \leq 0.90$ (180 files, excluded)
        \item \textbf{UNCERTAIN}: Conflicting signals (1{,}067 files, excluded)
    \end{itemize}
    \item Only STRONG\_VM and STRONG\_NVM files were used for training (24{,}812 total).
\end{enumerate}

This pseudo-labeling approach achieves a weighted accuracy of $\sim$90\% on the training set, which is sufficient for training a model that generalizes to 99.3\% on independently hand-labeled data.

\end{document}